# Large-Size Free-Standing Single-crystal β-Ga$_2$O$_3$ Membranes Fabricated by Hydrogen Implantation and Lift-Off

Yixiong Zheng[1], Zixuan Feng[2], A F M Anhar Uddin Bhuiyan[2], Lingyu Meng[2], Samyak Dhole[1], Quanxi Jia[1], Hongping Zhao[2,3], Jung-Hun Seo[1, a)]

[1]Department of Materials Design and Innovation, University at Buffalo, The State University of New York, Buffalo, NY USA 14260

[2]Depertment of Electrical and Computer Engineering, The Ohio State University, Columbus, Ohio 43210, USA

[3]Department of Materials Science and Engineering, The Ohio State University, Columbus, Ohio 43210, USA

[a)]Author to whom correspondence should be addressed: junghuns@buffalo.edu

## Abstract

In this paper, we have demonstrated the large-size free-standing single-crystal β-Ga$_2$O$_3$ NMs fabricated by the hydrogen implantation and lift-off process directly from MOCVD grown β-Ga$_2$O$_3$ epifilms on native substrates. The optimum implantation conditions were simulated with a Monte-Carlo simulation to obtain the high hydrogen concentration with a narrow ion distribution at the desired depth. Two as grown β-Ga$_2$O$_3$ samples with different orientation ([100] and [001]) were used and successfully create 1.2 μm thick β-Ga$_2$O$_3$ NMs without any physical damages. These β-Ga$_2$O$_3$ NMs were then transfer-printed onto rigid and flexible substrates such as SiC substrate and polyimide substrate. Various material characterizations were performed to investigate the crystal quality, surface morphology, optical property, mechanical property, and bandgap before and after the lift-off and revealed that good material quality is maintained. This result offers several benefits in that the thickness, doping, and size of β-Ga$_2$O$_3$ NMs can be fully controlled. Moreover, more advanced β-Ga$_2$O$_3$-based NM structures such as (Al$_x$Ga$_{1-x}$)$_2$O$_3$/Ga$_2$O$_3$ heterostructure NMs can be directly created from their bulk epitaxy substrates thus this result provides a viable route for the realization of high performance β-Ga$_2$O$_3$ NM-based electronics and optoelectronics that can be built on various substrates and platforms.

Beta-phase gallium oxide (β-Ga$_2$O$_3$) has attracted much attention as a promising wide bandgap semiconductor candidate due to its large bandgap with a high breakdown field and decent electron mobility.[1-4] The availability of large-size high-quality single-crystalline β-Ga$_2$O$_3$ native substrate and epitaxy layer via well-known substrate growth techniques such as the Czochralski method[5] and the Float Zone method[6] or by thin-film growth techniques such as molecular beam epitaxy (MBE)[7], metal-organic chemical vapor deposition (MOCVD)[8, 9], and halide vapor phase epitaxy (HVPE)[10] enable various large-scale high-performance electronics and optoelectronic applications.[11-14] Besides these advantageous material properties, β-Ga$_2$O$_3$ can also be mechanically exfoliated into a thin layer of β-Ga$_2$O$_3$ due to their different bonding strengths among different crystal orientations that associated with a monoclinic crystal structure. Thus, thin single-crystal β-Ga$_2$O$_3$ layers, also called β-Ga$_2$O$_3$ nanomembranes (NMs), can be produced directly from the bulk β-Ga$_2$O$_3$ substrate by a mechanical exfoliation method.[15-18] A free-standing form of β-Ga$_2$O$_3$ NM offers a new route to realize unique structures or device applications because it can be simply transfer-printed onto any desired substrates and form unique heterostructures or novel flexible electronics.[19, 20] For example, heterogeneous integration of transfer-printed β-Ga$_2$O$_3$ NMs onto a diamond substrate can be used to dissipate heat from β-Ga$_2$O$_3$ to compensate for the poor thermal property of β-Ga$_2$O$_3$ or novel heterojunctions.[21-23] Transfer-printed β-Ga$_2$O$_3$ NMs have also been used to realize flexible high-power switching devices or flexible solar-blind photodetectors that exhibited a comparable performance with that of their bulk counterparts.[24-26] However, one critical drawback is that β-Ga$_2$O$_3$ NMs are always cleaved at 77º angle to [201] direction when it is mechanically exfoliated due to the monoclinic crystal structure of β-Ga$_2$O$_3$ with a 103º angle. Therefore, the shape of the exfoliated β-Ga$_2$O$_3$ NM is always narrow and long. Our recent study also revealed that a wider β-Ga$_2$O$_3$ NMs leads to a thicker β-Ga$_2$O$_3$ NM.[26] For example, a 10 nm- and a 600 nm- thick β-Ga$_2$O$_3$ NMs typically show a width of 5 ~ 6 μm and 20 ~ 30 μm, respectively. In addition, it is difficult to precisely control a thickness of β-Ga$_2$O$_3$ NMs because the thickness can only be roughly controlled by the number of mechanical exfoliation steps. Another issue also related to the cleavage angle of β-Ga$_2$O$_3$ is that it is nearly impossible to create β-Ga$_2$O$_3$ NMs that have vertically grown multi-epitaxy layers, because all β-Ga$_2$O$_3$ NMs are exfoliated at a certain angle from the surface; thus severely limits the creation of functional β-Ga$_2$O$_3$ NM such as (Al$_x$Ga$_{1-x}$)$_2$O$_3$/Ga$_2$O$_3$ heterostructure NMs for various advanced electronics and optoelectronic applications.

In this paper, we demonstrated a large-size free-standing single-crystal β-Ga$_2$O$_3$ NMs fabricated by hydrogen implantation and lift-off directly from MOCVD grown epitaxy β-Ga$_2$O$_3$ samples. The use of hydrogen ions not only minimizes the crystal damage during ion implantation due to their light and small volume but also enables us to effectively separate the top portion of β-Ga$_2$O$_3$ from the substrate. Different implantation conditions were simulated with a Monte-Carlo simulation using a Silvaco Victory 2D Process simulator to obtain the high hydrogen concentration with a narrow ion distribution at a desired depth. Two β-Ga$_2$O$_3$ source wafers that were grown on different crystal orientations ([100] and [001]) were used and successfully create 1.2 μm thick β-$Ga_2O_3$ NMs without any physical damages. These β-$Ga_2O_3$ NMs were then transfer-printed onto rigid and flexible substrates such as SiC substrate and polyimide substrate. Various material characterizations were performed to investigate the crystal quality, surface morphology, optical property, mechanical property, and bandgap before and after the separation and revealed that no noticeable differences were observed. This result offers several benefits in that the thickness and size of β-$Ga_2O_3$ NMs can be accurately controlled as opposed to the narrow stripe-shaped β-$Ga_2O_3$ NMs from the uncontrolled conventional mechanical exfoliation method. Moreover, β-$Ga_2O_3$ NM-based functional free-standing semiconductor NMs such as $(Al_xGa_{1-x})_2O_3/Ga_2O_3$ heterostructure NMs can be directly created from their original hetero-epitaxy wafers without being restricted by their dimensional factors and easily integrated with any platform. Thus, this result provides a viable route to high performance β-$Ga_2O_3$ NM-based electronics and optoelectronics that can be built on various substrates such as flexible plastic or metallic substrates or different semiconductor platforms.

**Figure 1(a)** shows the schematic illustration of the β-$Ga_2O_3$ NM separation process. In this experiment, two 500 μm thick Fe doped β-$Ga_2O_3$ substrates (Novel Crystal Technology Inc.) with different orientations ([100] and [001]) and a Fe concentration of $1 \times 10^{19}$ cm$^{-3}$ were used. On top of these substrates, a 200 nm thick unintentionally doped homoepitaxy β-$Ga_2O_3$ thin-film grown by metalorganic chemical vapor deposition (MOCVD) as depicted in **Figure 1(a)(i)**. The detail of the samples can be found in **Figure S1** of Supplementary Information. The separation process starts with the hydrogen ion implantation (**Figure 1(a)(i)-(ii)**). Prior to the ion implantation, both samples were thoroughly cleaned with acetone, isopropyl alcohol, and deionized water with mild sonification. The detail of the ion implantation condition will be discussed in below related to **Figure 3**. In this experiment, the hydrogen implantation was performed with an energy of 190 keV and a dose of $2 \times 10^{16}$ cm$^2$ to produce the dense hydrogen layer 1.2 μm underneath the wafer surface. The sample was then annealed at 250 °C for 24 hours to activate implanted hydrogen ions (**Figure 1(a)(iii)**). As implanted hydrogen ions reacted and became hydrogen gas (hydrogen micro-bubbles), the top portion of β-$Ga_2O_3$ was gradually separated from the substrate. This thin β-$Ga_2O_3$ can be now called β-$Ga_2O_3$ NM and gently registered on the β-$Ga_2O_3$ substrate without moving as shown in **Figure 1(a)(iv)**. Then, as depicted in **Figure 1(a)(v)-(viii)**, β-$Ga_2O_3$ NMs were transfer-printed onto rigid and flexible substrates using an elastomeric stamp (poly(dimethylsiloxane) (PDMS)) to construct the final structure, namely, β-$Ga_2O_3$ NMs on a foreign substrate (**Figure 1(a)(viii)**). The detail of the transfer printing process can be found elsewhere[27, 28]. In this experiment, the SiC and the polyimide substrates were used to represent rigid and flexible form of foreign substrate, but β-$Ga_2O_3$ NMs can be transfer-printed onto any type of rigid or flexible substrates. **Figure 1(b)** shows multiple images of 15 mm × 10 mm size single piece β-$Ga_2O_3$ NM on a PDMS stamp with three different magnifications to show the detail of the lifted-off β-$Ga_2O_3$ NM. **Figure 2(a)** shows a three-dimensional surface profile of β-$Ga_2O_3$ NM on the PDMS surface taken after lifting up using Profilm3D Filmetric surface profiler. The surface profile image scanned ~500 μm² area with a spatial resolution of < 5 nm. **Figure 2(b)** shows the two-dimensional depth profile between the "A" and "B" points of **Figure 2(a)**, indicating that the surface of β-Ga₂O₃ NM is smooth and uniform. **Figure 2(c)** shows that the thickness of β-Ga₂O₃ NMs was measured to be 1.2 μm, and the surface roughness of 1.8 nm which is the same as the surface roughness of the β-Ga₂O₃ epifilms before the lift-off. The angled scanning electron microscopy (SEM) images (**Figure 2(c)**) also show that the surface is smooth without any cracks or fractures. These imaging results confirm that the layer separation occurred exactly at the depth as originally designed.

The process design for hydrogen implantation is a critical and most important step that determines a thickness of β-Ga₂O₃ NM after the lift-off process. The implantation modeling was performed using an ion implantation module of the Silvaco Victory 2D Process simulator to accurately predict the hydrogen distribution in the bulk β-Ga₂O₃ substrate. The hydrogen profile was modeled based on a Monte-Carlo method with 32000 hydrogen ions using an implantation energy ranging from 70 keV to 300 keV with a fixed dose at $2 \times 10^{16}$ cm⁻². In each hydrogen implantation profile, we captured the depth where the hydrogen concentration exceeds mid-$10^{20}$ cm⁻³, because the layer separation will occur at this depth, thus the thickness of β-Ga₂O₃ NM after the lift-off process can be estimated. **Figure 3(a)** shows the summary data showing a predicted thickness of the separated β-Ga₂O₃ NMs as a function of different implantation energy with a fixed dose of $2 \times 10^{16}$ cm⁻². A linear relationship between β-Ga₂O₃ NM thickness and implantation energy was observed as predicted by the energy (E) and projected range (R) relationship in the ion implantation theory: $R = \frac{1}{N}\int_0^E dE/S(E)$, where N is the number of ions per unit volume.[29] Therefore, it is possible to accurately control the separation depth by controlling the implantation energy. In fact, this hydrogen implantation and layer-separation process is similar to the well-known SMART-CUT process, in that a dense hydrogen layer is used to separate the layer from the bulk substrate to fabricate various XOI (Semiconductor X-On Insulator) wafers such as SOI, GeOI, SiCOI and etc. [20, 30-32] Thus this hydrogen implantation and layer-separation process is a highly reliable process. **Figure 3(b)** shows the modeled hydrogen profile that is used in this experiment. According to this simulation result, the hydrogen implantation process with the energy of 190 keV and a dose of $2 \times 10^{16}$ cm² yield a 1 μm thick β-Ga₂O₃ NM. The excessively high energy and dose level compared with the lift-off process for other materials are responsible for the high material density of β-Ga₂O₃.[30, 31] It should be noted that the hydrogen concentration at the separation region needs to be greater than $1 \times 10^{20}$ cm⁻³ to create microbubbles for the layer lift-off without any cracks or fractures, because the failure to achieve dense hydrogen bubbles results in the partial layer separation and causes defects and cracks in β-Ga₂O₃ NMs. The star mark in **Figure 3(a) and (b)** indicate the actual thickness of β-Ga₂O₃ NM (~1.2 μm) after the lift-off process which agrees well with the predicted value. In this experiment, we designed a 1.2 μm thick β-Ga₂O₃ NM for the proof of concept, but a wide range of β-Ga$_2$O$_3$ NM thicknesses (from a few hundreds of nm to several tens of μm range) can be realized by changing the hydrogen implantation condition.

Various material characterizations were exhibited to investigate differences before and after the layer lift-off process. In this characterization, β-Ga$_2$O$_3$ NM on the PDMS stamp which corresponds to the step (vi) in **Figure 1(a)** was used to avoid any possible material damages by the transfer-printing process that can affect the result. First, X-ray diffraction (XRD) measurements were performed to investigate the crystalline quality of lifted-off β-Ga$_2$O$_3$ NM using a PANalytical Empyrean X-ray diffractometer (Cu Kα radiation with operating voltage/current of 45 kV/40 mA) under ambient conditions with a 20–60° 2θ scattering angle range. As shown in **Figure 4(a) and (b)**, each XRD scan clearly indicates the peaks of [400], [600], and [800] planes for β-Ga$_2$O$_3$ NM from the [100]-oriented β-Ga$_2$O$_3$ substrate and [002] plane for β-Ga$_2$O$_3$ NM from the [001]-oriented β-Ga$_2$O$_3$ substrate. The full width at half maximum (FWHM) values of the most dominant XRD profiles from each sample, namely, [200] plane for the [100]-oriented β-Ga$_2$O$_3$ NM and [002] plane for the β-Ga$_2$O$_3$ NM, are measured to be 0.013° and 0.029°. (see **Figure S2** in Supplementary Information) These FWHM values and 2θ remain unchanged compared to their bulk counters (**Figure S3** in Supplementary Information) and it implies that the quality of lifted-off β-Ga$_2$O$_3$ NMs remain unchanged. We also compared the Raman spectrum from a bulk β-Ga$_2$O$_3$ to that of lifted-off β-Ga$_2$O$_3$ NMs using a Renishaw InVia Raman spectroscopy that was equipped with a 514 nm green laser and a ×50 objective lens. **Figure 4(c) and (d)** show the Raman spectra that were taken from the bulk β-Ga$_2$O$_3$ and lifted-off β-Ga$_2$O$_3$ NMs. The Raman intensity from the bulk β-Ga$_2$O$_3$ is much stronger than that of lifted-off β-Ga$_2$O$_3$ NM due to the difference in their physical thicknesses (500 μm vs. 1.2 μm). All Raman spectra present 11 typical Raman modes from 100 cm$^{-1}$ to 900 cm$^{-1}$ without any noticeable shifting indicating that the β-Ga$_2$O$_3$ NMs do not have any damage in the crystal structure and internal strain after the lift-off process compared to bulk β-Ga$_2$O$_3$. However, [100] β-Ga$_2$O$_3$ (both bulk and NM) has a stronger A$_{g,3}$ Raman mode (appeared at 200 cm$^{-1}$) and weaker A$_{g,10}$ Raman mode (appeared at 760 cm$^{-1}$) which indicates that the vibration modes of the Ga–O chain in the Ga$_I$O$_4$ tetrahedron have a dominant position in [100] β-Ga$_2$O$_3$ compared with [001] β-Ga$_2$O$_3$. Interestingly, the A$_{g,10}$ Raman mode becomes significantly weaker in both [100] and [001] β-Ga$_2$O$_3$ NMs. The A$_{g,10}$ Raman mode is ascribed to the combination of the symmetrical stretching vibration of the Ga$_I$(O$_I$-O$_{III}$) bond of the Ga$_I$O$_4$ unit and the bending vibration of the Ga$_I$(O$_{II}$)$_2$ bond and this mode is affected by the adjacent octahedron.[33] The Fe dopants in bulk β-Ga$_2$O$_3$ typically replace GaII atoms in the center of Ga$_{II}$O$_6$ octahedron, therefore the A$_{g,10}$ Raman mode in bulk β-Ga$_2$O$_3$ tends to be very strong. On the contrary, undoped β-Ga$_2$O$_3$ has a weaker A$_{g,10}$ Raman mode as a result of lower dopant concentrations.[34] In β-Ga$_2$O$_3$ NMs, the relative thickness of Fe doped β-Ga$_2$O$_3$ in β-Ga$_2$O$_3$ NM is reduced from 500 μm to 0.9 μm after the lift-off process. Therefore, the relative A$_{g,10}$ Raman mode vibration is also noticeably reduced both in [100] and [001] β-Ga$_2$O$_3$ NMs. **Figure 5** shows Raman spectra taken after transfer-printing [001]-oriented β-Ga$_2$O$_3$ NMs onto a SiC and polyimide substrates. As shown in **Figure 5(a)**, the β-Ga$_2$O$_3$ NM on SiC substrate clearly presents 11 β-Ga$_2$O$_3$ characteristic modes as well as SiC peaks at 780 nm and 980 nm respectively without any peak shifting, suggesting that β-Ga$_2$O$_3$ NM does not suffer from any residual stress. Also, the similar phenomenon was measured from β-Ga$_2$O$_3$ NM on polyimide structure that no noticeable peak shifting was observed as shown in **Figure 5(b)**. Therefore, the lifted β-Ga$_2$O$_3$ NMs did not experience any strain or material degradation during the lift-off process and the integration on the foreign substrate. After we confirmed the crystal quality of the transferred β-Ga$_2$O$_3$ NMs, we performed a strain-Raman spectral study to investigate Raman shifts under different uniaxial strain conditions. In order to accurately measure the changes in the Raman spectrum under the uniaxial strain condition, we employed convex and concave molds that have different curve radii ranging from 110 mm to 20 mm which corresponds to the uniaxial strains up to 0.32 % of the tensile strain (for the convex mold) and up to 0.19 % of the compressive strain (for the concave mold). The most dominant peak A$_{g,3}$ peak at 200.4 cm$^{-1}$ was chosen to trace the strain-dependent characteristics. As shown in **Figure 5(c), Figure S4, and Figure S5**, the peak shifting value of A$_{g,3}$ was measured to be 2.56 cm$^{-1}$ /Δε. This value is larger than typical 2D transition metal dichalcogenide (TMD) semiconductors[35], but similar to other single crystal semiconductor NMs[36].

To further investigate the impact of the hydrogen implantation and lift-off process on the optical property and bandgap of β-Ga$_2$O$_3$ NM, the optical property characterization was performed. Firstly, the refractive index (*n*) and extinction coefficient (*k*) were measured at the wavelength from 200 nm to 400 nm using a customized UV-to-visible spectrometer. As shown in **Figure 6(a) and (b)**, *n* and *k* values for bulk β-Ga$_2$O$_3$ and β-Ga$_2$O$_3$NMs of both [100] and [001] orientations on SiC substrates were compared. Both n and k values are very similar between the bulk format and NM format of β-Ga$_2$O$_3$. The small difference is probably attributed to the light reflection from the different substrates (β-Ga$_2$O$_3$ substrate vs. SiC substrate) and the slight difference in doping concentration of bulk β-Ga$_2$O$_3$ and β-Ga$_2$O$_3$ NM. Using these measured n and k values, the absorption coefficient can be obtained from the extinction coefficient using the following equation: $α(λ)=[2π·k(λ)]/λ$ by the complex index of refraction ($N = n - ik$) relationship.[37] Then, the optical bandgap for the direct electron transition can be calculated using the Tauc plot formula: $α·h·v=C(h·v-E_g)^{(1/2)}$, where $h·v$ is the photon energy, $λ$ is the wavelength and *C* is a constant.[37] From the Tauc plot, the bandgap ($E_g$) could be estimated by extrapolating the linear sections to the axis of energy ($h·v$). **Figure 6(c)** shows the estimated bandgap of bulk β-Ga$_2$O$_3$ and β-Ga$_2$O$_3$NMs of both [100] and [001] orientations. Interestingly, while the bandgap values of bulk [100] and [001] β-Ga$_2$O$_3$ are calculated to be 4.94 eV and 4.93 eV, the bandgap values of [100] and [001] β-Ga$_2$O$_3$ NMs are slightly decreased to be 4.94 eV and 4.88 eV, respectively. Although the difference in bandgap between the bulk β-Ga$_2$O$_3$ substrate and β-Ga$_2$O$_3$NM is about 0.2~0.3 eV, this difference can be explained by the total amount of dopants in bulk β-Ga$_2$O$_3$ substrate β-Ga$_2$O$_3$ NMs. Rafique et al. reported that a 100 times difference in doping concentration (~10$^{17}$ cm$^{-3}$ vs

~$10^{19}$ cm$^{-3}$) can result in ~1 % difference in bandgap of β-Ga$_2$O$_3$.[38] In fact, a few meV bandgap reduction by decreasing doping concentration is commonly observed in a similar material system such as GaAs and GaN.[39, 40] In our case, as depicted in **Figure 1(a)**, a 500 μm thick β-Ga$_2$O$_3$ substrate contains Fe dopants with a concentration of $1 \times 10^{19}$ cm$^{-3}$, while the top 200 nm epitaxy layer has an unintentional doping concentration of $< 1 \times 10^{16}$ cm$^{-3}$. When the bandgap of bulk β-Ga$_2$O$_3$ is measured, the effect on optical parameters by a 200 nm thick UID epitaxy β-Ga$_2$O$_3$ NM is almost negligible compared with that of 500 μm thick Fe-doped β-Ga$_2$O$_3$ substrate. However, when the optical parameters of β-Ga$_2$O$_3$ NM are measured, the 200 nm thick UID epitaxy β-Ga$_2$O$_3$ NM cannot be negligible, because the thickness of Fe-doped β-Ga$_2$O$_3$ is estimated to be ~900 nm. Therefore, the difference in bandgap is not by the hydrogen implantation and lift-off process, but by the relative thickness ratio between the Fe-doped bulk β-Ga$_2$O$_3$ and β-Ga$_2$O$_3$ NM, because the effect of the dopants becomes more dominant in β-Ga$_2$O$_3$ NMs

In conclusion, we have successfully demonstrated the large-size free-standing single-crystal β-Ga$_2$O$_3$ NMs fabricated by the hydrogen implantation and lift-off process directly from MOCVD grown β-Ga$_2$O$_3$ substrates. The optimum implantation conditions were simulated with a Monte-Carlo method to obtain the high hydrogen concentration with a narrow ion distribution at the desired depth. Two β-Ga$_2$O$_3$ source wafers with different orientations (i.e., [100]- and [001]-oriented β-Ga$_2$O$_3$ substrates) were used and successfully create 1.2 μm thick β-Ga$_2$O$_3$ NMs without any physical damages. These β-Ga$_2$O$_3$ NMs were then transfer-printed onto rigid and flexible substrates such as SiC substrate and polyimide substrate. Various material characterizations were performed to investigate the crystal quality, surface morphology, optical property, mechanical property, and bandgap before and after the separation and revealed that good material quality was maintained. This result offers several benefits in that the thickness and size of β-Ga$_2$O$_3$ NMs can be fully controlled. Moreover, more advanced β-Ga$_2$O$_3$ NM structures such as (Al$_x$Ga$_{1-x}$)$_2$O$_3$/Ga$_2$O$_3$ heterostructure NMs can be directly created from their bulk epitaxy wafers. Several following studies need to be conducted such as the impact of growth conditions such as growth temperature and n-type/p-type doping concentration on the surface roughness, defect density of the bulk source β-Ga$_2$O$_3$ to create the uniform and high-quality β-Ga$_2$O$_3$ NMs. Nevertheless, this result provides a viable route for the realization of high-performance β-Ga$_2$O$_3$ NM-based electronics and optoelectronics that can be built on various substrates and platforms.

ACKNOWLEDGEMENT

This work was supported by the National Science Foundation (Grant number: ECCS - 1809077) and partially by the seed grant by Research and Education in energy, Environment, and Water (RENEW) Institute at the University at Buffalo, the Air Force Office of Scientific Research FA9550-18-1-0479 (AFOSR, Dr. Ali Sayir), and the National Science Foundation (Grant number: ECCS – 1810041 and 2019753).

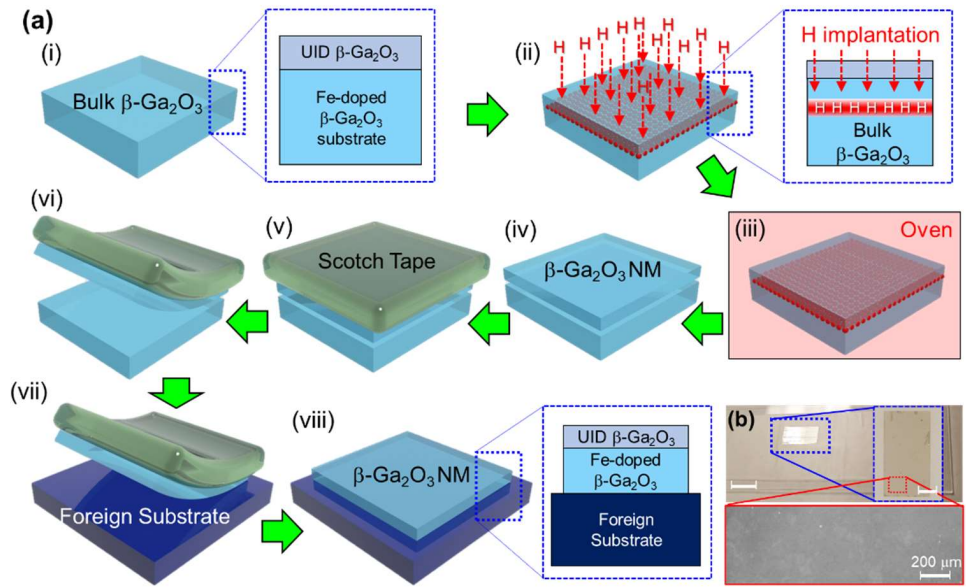

**Figure 1. (a)** a schematic illustration of the β-Ga₂O₃ lift-off process, **(b)** images of lifted β-Ga₂O₃ in different magnifications (upper left) x2, (upper right) x5, (bottom) x 400. Scale bars in upper left and right images are 10mm and 5mm respectively.

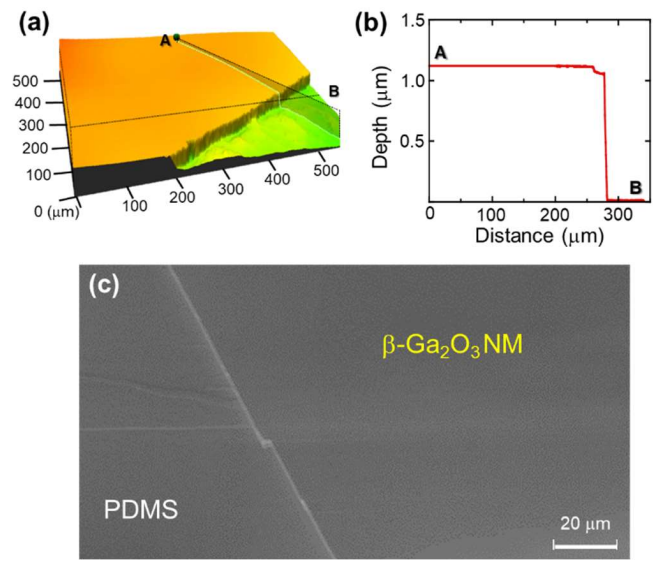

**Figure 2. (a)** three dimensional surface morphology of β-Ga₂O₃ NM and **(b)** two-dimensional depth profile between the point A and B in Figure 2(a), **(c)** an angled SEM image of β-Ga₂O₃ NM taken on a PDMS stamp.

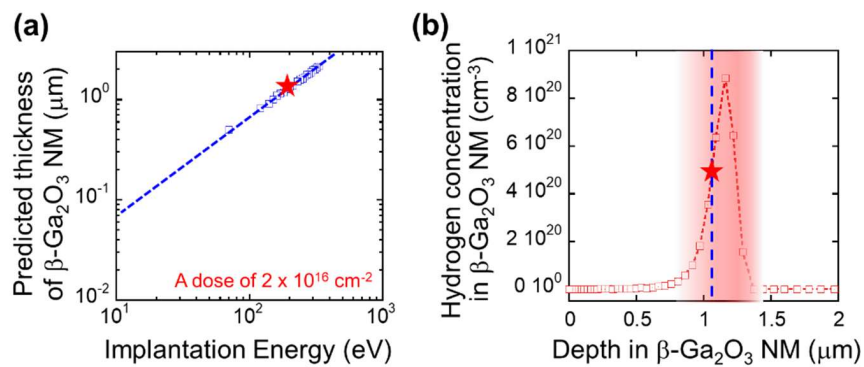

**Figure 3. (a)** a predicted thickness of β-Ga$_2$O$_3$ NM with respect to various implantation energy from 70 eV to 300 eV. **(b)** a simulated hydrogen concentration profile in β-Ga$_2$O$_3$ NM. A star mark in each figure denotes the actual thickness of β-Ga$_2$O$_3$ NM.

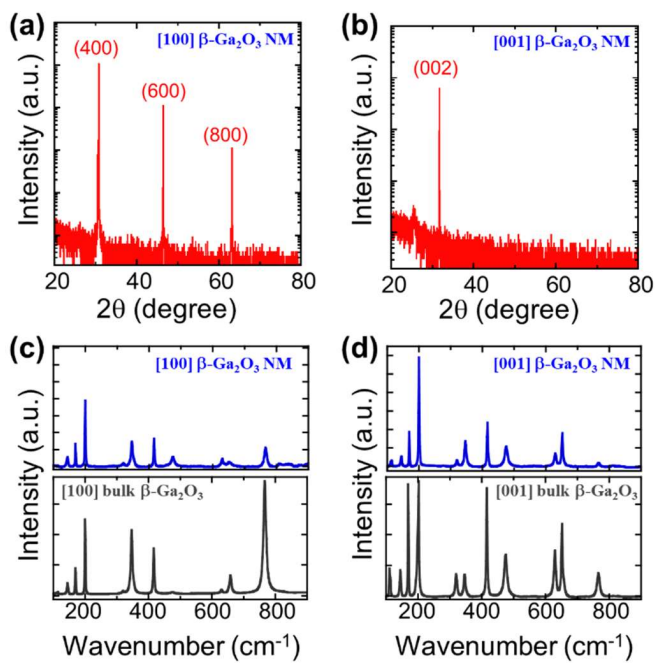

**Figure 4.** Measured XRD spectrum from **(a)** [100]-oriented β-Ga$_2$O$_3$ NM and **(b)** [001]-oriented β-Ga$_2$O$_3$ NM. Measured Raman spectrum from **(c)** [100]-oriented β-Ga$_2$O$_3$ NM and **(d)** [001]-oriented β-Ga$_2$O$_3$ NM compared with their bulk form of substrates (lower panel of Figure 4(c) and (d)).

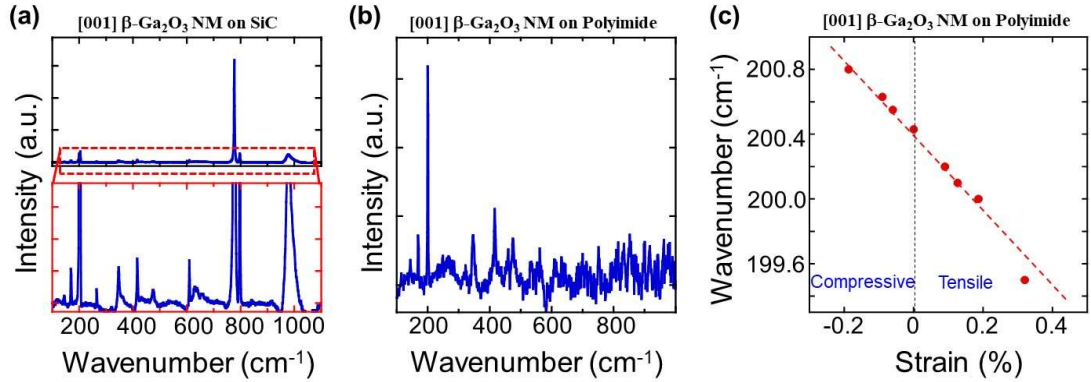

**Figure 5.** Measured Raman spectrum from [001]-oriented β-Ga$_2$O$_3$ NM transferred on SiC substrate. A bottom section of the plot shows the zoomed-in view of the Raman spectrum and **(b)** Measured Raman spectrum from [001]-oriented β-Ga$_2$O$_3$ NM transferred on on polyimide substrate. **(c)** Measured A$_{g,3}$ Raman modes as a function of applied strain from 0.19% of the tensile strain to 0.32 % of the compressive strain.

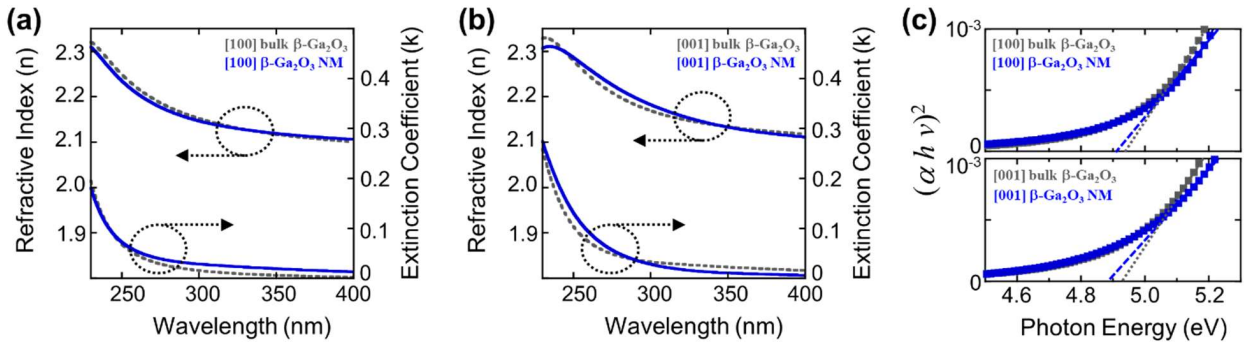

**Figure 6.** Measured optical properties: (left axis) refractive index (*n*) and (right axis) extinction coefficient (*k*) of **(a)** [100]-oriented β-Ga$_2$O$_3$ NM and **(b)** [001]-oriented β-Ga$_2$O$_3$ NM compared with their bulk form of substrates. **(c)** extracted bandgap of (upper) [100]-oriented β-Ga$_2$O$_3$ NM and (lower) [001]-oriented β-Ga$_2$O$_3$ NM compared with their bulk form of substrates.